\documentclass[a4paper]{jpconf}
\usepackage{graphicx}
\newcommand{\solm}{M$_{\odot}$\ }

\begin{document}

\title{Coordinated mm/sub-mm observations of 
\\
Sagittarius A* in May 2007}

\author{
       D Kunneriath$^{1,2}$,
       A Eckart$^{1,2}$,
       S Vogel$^{3}$,   
       L Sjouwerman$^{4}$,
       H Wiesemeyer$^{5}$,  
       R Sch\"odel$^{6}$,
       F K. Baganoff$^7$,
       M Morris$^8$,
       T Bertram$^1$,
       M Dovciak$^{9}$, 
       D Downes$^{10}$, 
       W J Duschl$^{11,12}$,
       V Karas$^{9}$, 
       S K\"onig$^{1,2}$,
       T Krichbaum$^{2}$,
       M Krips$^{13}$,
       R-S Lu$^{2,1}$,
       S Markoff$^{14}$,
       J Mauerhan$^8$,
       L Meyer$^8$,
       J Moultaka$^{15}$,
       K Muzic$^{1,2}$,
       F Najarro$^{16}$,
       K Schuster$^{10}$,
       C Straubmeier$^1$,
       C Thum$^{10}$,
       G Witzel$^1$,
       M Zamaninasab$^{1,2}$ and
       A Zensus$^{2}$
}

\address{
$^1$  University of Cologne, Z\"ulpicher Str. 77, D-50937 Cologne, Germany\\
$^2$  Max-Planck-Institut f\"ur Radioastronomie, Auf dem H\"ugel 69, 
    53121 Bonn, Germany\\
$^{3}$  Department of Astronomy, University of Maryland, College Park, 
    MD 20742-2421, USA\\
$^{4}$ National Radio Astronomy Observatory,
       PO Box 0, Socorro, NM 87801, USA\\
$^{5}$  IRAM, Avenida Divina Pastora, 7, N\'ucleo Central, 
      E-18012 Granada, Spain\\
$^6$  Instituto de Astrof\'isica de Andaluc\'ia, Camino Bajo de
    Hu\'etor 50, 18008 Granada, Spain \\
$^7$ Center for Space Research, Massachusetts Institute of
            Technology, Cambridge, MA~02139-4307, USA \\
$^8$  Department of Physics and Astronomy, University of California, 
     Los Angeles, CA 90095-1547, USA\\
$^9$ Astronomical Institute, Academy of Sciences, 
        Bo\v{c}n\'{i} II, CZ-14131 Prague, Czech Republic \\
$^{10}$ Institut de Radio Astronomie Millimetrique, Domaine Universitaire, 
    38406 St. Martin d'Heres, France\\
$^{11}$ Institut f\"ur Theoretische Physik und Astrophysik,
        Christian-Albrechts-Universit\"at zu Kiel, Leibnizstr. 15
        24118 Kiel, Germany \\
$^{12}$ Steward Observatory, The University of Arizona, 933 N. 
     Cherry Ave. Tucson, AZ 85721, USA\\
$^{13}$ Harvard-Smithsonian Center for Astrophysics, SMA project, 
     60 Garden Street, MS 78 Cambridge, MA 02138, USA\\
$^{14}$    Astronomical Institute `Anton Pannekoek', 
        University of Amsterdam, Kruislaan 403,
        1098SJ Amsterdam, the Netherlands\\
$^{15}$ Observatoire Midi-Pyr\'en\'ees,
        14, Avenue Edouard Belin, 31400 Toulouse, France\\
$^{16}$ Instituto de Estructura de la Materia, 
        Consejo Superior de Investigaciones Cientificas, 
        CSIC, Serrano 121, 28006 Madrid, Spain\\
            }

\ead{eckart@ph1.uni-koeln.de}

\begin{abstract}
At the center of the Milky Way, with a distance of $\sim$8~kpc, the
compact source Sagittarius~A* (SgrA*) can be associated with a super
massive black hole of $\sim$4$\times$10$^6$\solm.  SgrA* shows strong
variability from the radio to the X-ray wavelength domains.  Here we
report on simultaneous NIR/sub-millimeter/X-ray observations from May
2007 that involved the NACO adaptive optics (AO) instrument at the
European Southern Observatory's Very Large Telescope, the Australian
Telescope Compact Array (ATCA), the US mm-array CARMA, the IRAM~30m
mm-telescope, and other telescopes.  We concentrate on the time series
of mm/sub-mm data from CARMA, ATCA, and the MAMBO bolometer at the
IRAM 30m telescope.

\end{abstract}

\section{Introduction}

The investigation of the dynamics of stars has provided compelling
evidence for the existence of a super-massive black hole (SMBH) at the
center of the Milky Way.  At a distance of only $\sim$8~kpc a SMBH of
$\sim$4$\times$10$^6$\solm can convincingly be identified with
the compact radio and infrared source Sagittarius A* (Sgr~A*) (Eckart
\& Genzel 1996, Genzel et al.\ 1997, 2000, Ghez et al.\ 1998, 2000,
2004ab, 2005, Eckart et al.\ 2002 , Sch\"odel et al.\ 2002, 2003,
Eisenhauer et al.\ 2003, 2005).  Due to its proximity Sgr~A* provides
us with a unique opportunity to understand the physics of super
massive black holes at the nuclei of galaxies.

Studies of the variable polarized NIR emission and simultaneous
radio/NIR/X-ray observations of SgrA* are ideally suited to obtain
deep insights into the relativistic physics within 10-100
Schwarzschild radii of the SMBH associated
with SgrA*.  In the following,  we assume for Sgr~A*
$R_s$=2$R_g$=2GM/c$^2$$\sim$8~$\mu$as, with $R_s$ being one
Schwarzschild radius and $R_g$ the gravitational radius of the SMBH.

Sgr~A* is remarkably faint in all wavebands, challenging current
theories of matter accretion and radiation surrounding SMBHs. The
feeble emission ($<10^{-9}$ of the Eddington rate) is due to a
combination of a low accretion rate with a low radiation efficiency.
Theoretical interpretations at present focus on radiatively
inefficient accretion flow and jet models.  For a recent summary of
accretion models and variable accretion of stellar winds onto Sgr A*
see Yuan (2006) and Cuadra \& Nayakshin (2006) and the references in
these papers.

The first successful simultaneous NIR/X-ray campaigns combined NACO
and Chandra as well as mostly quasi-simultaneous mm-data from BIMA,
SMA, and VLA (Eckart et al.\ 2004, 2006a).  The NIR/X-ray variability is
probably also linked to the variability at radio through
sub-millimeter wavelengths showing that variations occur on time
scales from hours to years (Bower et al.\ 2002, Herrnstein et al.\ 2004,
Zhao et al.\ 2003, 2004, Mauerhan 2005). In various observing
campaigns we found simultaneous NIR/X-ray flare variations 
(Eckart et al.\ 2008, 2006, 2005, 2004a, Marrone et al.\ 2008, 
Yuesf-Zadeh et al.\ 2008), indications of significant,
  possibly quasi-periodic, sub-structure within NIR
flares (Genzel et al.\ 2003a, 2003b, Eckart et al.\ 2006, Ghez 2003b),
and highly polarized emission (Eckart et al.\ 2006).

The 10$^{33-34}$~erg/s flares can be explained with a synchrotron
self-Compton (SSC) model involving up-scattered sub-millimeter photons
from a compact source component (e.g. Eckart et al.\ 2004, Eckart et
al. 2006a).  Inverse Compton scattering of the THz-peaked flare
spectrum by the relativistic electrons then accounts for the X-ray
emission.  This model allows for NIR flux density contributions from
both the synchrotron and SSC mechanisms.

The NIR flare emission is polarized with a well defined range over
which the position angle of the polarized emission is changing
(60$^o$$\pm$20$^o$, Eckart et al.\ 2006b, Meyer et al.\ 2006ab, 2007).
All these observations can be explained in a model of a temporary
accretion disk harboring one or several bright orbiting spot(s),
possibly in conjunction with a short jet, and suggest a stable
orientation of the source geometry over the past few years.
Hawley \& Balbus (1991) and Balbus (2003) outline a model of an
expanding hot spot within an inclined temporary accretion disk.
However, the radio/sub-mm/NIR observations also indicate adiabatic
expansion within an SSC model (Eckart et al.\ 2005, 2006, 2008
Yusef-Zadeh et al.\ 2006ab, 2008, Marrone et al.\ 2008) and the emission
very likely originates from a combination of a temporal accretion disk
and a short, low-luminosity jet.

The millimeter/submillimeter wavelength polarization of Sgr A* is
variable in both magnitude and position angle on timescales down to a
few hours.  Marrone et al.\ (2007) present simultaneous observations
made with the Submillimeter Array (SMA) polarimeter at 230 and 350 GHz
with sufficient sensitivity to determine the polarization degree and
rotation measure within each band.  From their measurements they
deduce an accretion rate that does not vary by more than 25\% and -
depending on the equipartition constraints and the magnetic field
configuration - amounts to 2$\times$10$^{-5}$ to 2$\times$10$^{-7}$
\solm yr$^{-1}$.  The mean intrinsic position angle of the measured
polarization is 167$^\circ$$\pm$7$^\circ$ with variations of
$\sim$31$^\circ$ that must originate in the sub-millimeter photosphere
of SgrA*.

Here, we present mm- and sub-mm-data that have been obtained using
CARMA\footnote{Support for CARMA construction was derived from the
  states of California, Illinois, and Maryland, the Gordon and Betty
  Moore Foundation, the Kenneth T. and Eileen L. Norris Foundation,
  the Associates of the California Institute of Technology, and the
  National Science Foundation.  Ongoing CARMA development and
  operations are supported by the National Science Foundation under a
  cooperative agreement, and by the CARMA partner universities.},
ATCA\footnote{ATCA is operated by the Australia Telescope National
  Facility, a division of CSIRO, which also includes the ATNF
  Headquarters at Marsfield in Sydney, the Parkes Observatory and the
  Mopra Observatory near Coonabarabran.}  and the MAMBO bolometer at
the IRAM\footnote{The IRAM 30m millimeter telescope is operated by the
  Institute for Radioastronomy at millimeter wavelengths - Granada,
  Spain, and Grenoble, France.}~30m telescope during a coordinated,
multi wavelength observing campaign of SgrA* in May 2007.  .

\section{Observations and Data Reduction}

Interferometric observations in the mm/sub-mm wavelength 
domain are especially well suited to separate the flux density 
contribution of SgrA* from the thermal emission of the Circum Nuclear Disk 
(a ring-like structure of gas and dust surrounding the Galactic Center at a distance 
of about 1.5-4~pc, see, e.g., G{\"u}sten et al.\, 1987 or Christopher
et al.\, 2005).

\begin{figure}
\begin{center}
\includegraphics[width=16cm, angle=-0]{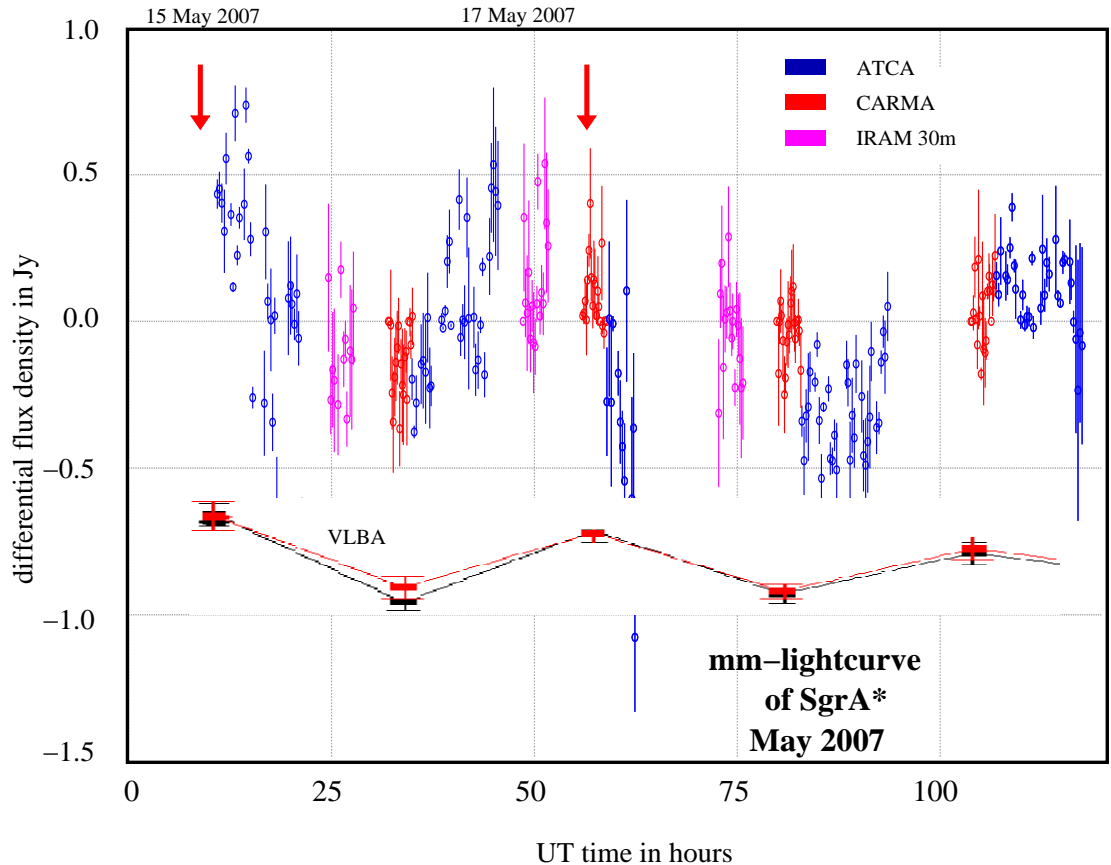}
\end{center}
\caption{ Combined differential light curve of SgrA* in the mm/sub-mm
  domain for the May 2007 observing run.  The MAMBO~2 bolometer  at
  the IRAM 30m-telescope was
  operated at a central wavelength of 1.2\,mm (250 GHz).
  The CARMA data were
  centered at 100~GHz and the ATCA data at 86~GHz.  We also show the
  daily 7mm flux density averages of our 2007 VLBA session 
 (black and red symbols represent the signals from the R and L circularly
 polarized channels).
   The red
  vertical arrows mark the peak times of NIR flares observed with
  NACO.  The time axis is labeled with UT hours starting at 00~h
  on May 15.
     }
\label{fig:1}
\end{figure}

In May 2007, global coordinated multi-wavelength observations were
carried out in the NIR and mm regimes to study the variability of Sgr
A*.  We observed the galactic center at 100 and 86~GHz (3 mm
wavelength) with the two mm-arrays CARMA and ATCA.  In addition
we observed with the MAMBO~2 bolometer at the IRAM 30\,m-telescope at a
wavelength of 1.3~mm.  CARMA (Combined Array for Research in mm-wave
Astronomy) is located in Cedar Flat, Eastern California, and consists
of 15 antennas (6 x 10.4~m and 9 x 6.1~m ).  The Australia Telescope
Compact Array (ATCA), at the Paul Wild Observatory, is an array of 6
22-m telescopes located in Australia, about 25 km west of the town of
Narrabri in rural NSW (about 500 km north-west of Sydney).  The
interferometer data were mapped using the {\it Miriad} interferometric data
reduction package.

The Max-Planck Millimeter Bolometer (MAMBO~2) array is installed at
the IRAM~30m telescope on Pico Veleta, Spain. The 37 channel array
of the precursor instrument MAMBO
has been successfully used by many observers since the end of
1998. Winter 2001/2002 was the first season of the new MAMBO-2 version
with 117 pixels.  The 37 channel MAMBO is used at the 30 m telescope
as a backup system now.  Both systems work at 1.2 mm wavelength and
have a He-3 fridge to operate the bolometers at a temperature of
300~mK.  The bolometer data was reduced using the bolometer array data
reduction, analysis, and handling software package, the BoA (Bolometer
Data Analysis).

\section{Preliminary results of the May 2007 mm-observations}

\begin{figure}
\begin{center}
\includegraphics[width=16cm, angle=-0]{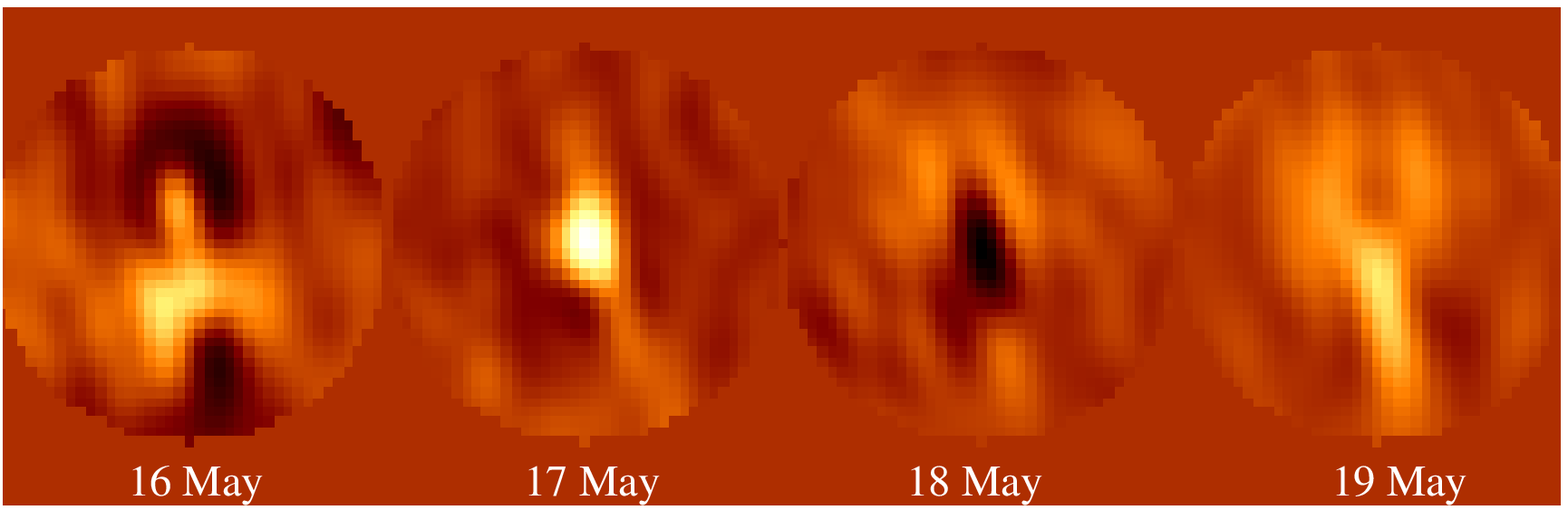}
\end{center}
\caption{Difference maps at 3\,mm of a 40$''$ diameter region centered 
on Sgr~A*, obtained from the difference between full synthesis maps of the
individual  days of CAMRMA observations and the full CARMA data set 
as described in the text. 
The figure shows that the flux density variations that are evident from
the differential light curves (see Fig.\,\ref{fig:1}) can also be seen in the maps constructed from the
corresponding data.
}
\label{fig:2}
\end{figure}

\begin{figure}
\begin{center}
\includegraphics[width=13cm, angle=-0]{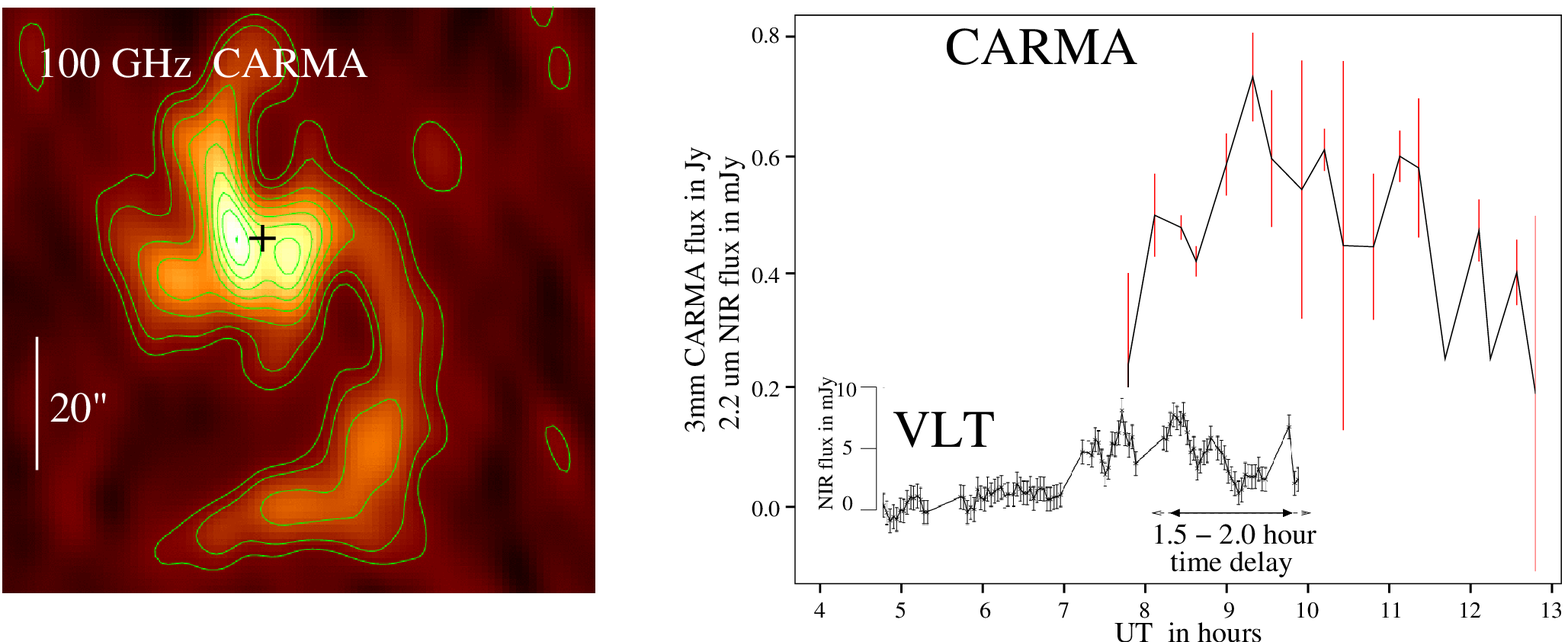}
\end{center}
\caption{{\bf Left:} Point source (i.e. SgrA*) subtracted Galactic Center CARMA 100~GHz map 
composed of the 3 best of 4 full D array coverages (0.05, 0.1, 0.2, ... Jy/beam).
The image shows the mini-spiral continuum and demonstrates the high data quality
of the 2-3$''$ resolution map.
{\bf Right:}
Zooming in on the May 17 CARMA and VLT data.
CARMA light curve of SgrA*  as derived from visibilities of the two independent
baselines between telescopes 3-1 and 14-6 (mean + differences as 1$^{st}$-order error est.). 
The {\bf inset} shows the NIR flux density obtained from the VLT at 2.2$\mu$m wavelength.
}
\label{fig:3}
\end{figure}

Based on the assigned VLT time we organized a large multi-frequency
campaign in May 2007 that included millimeter to MIR observations at
single telescopes and interferometers around the world.  The main
results are: 2 bright NIR flares (16~mJy and 5~mJy) and one CARMA 3mm
flare of 0.5~Jy.  A 10 day VLBI observing campaign performed with the
VLBA at 22, 43 and 86 GHz (1.3cm, 7mm, 3.4mm wavelength) is being
reduced. First sub-milliarcsecond resolution VLBI maps are presented
by Lu et al.\ (2008) in this edition.

In order to correct for extended flux contributions in the
interferometer data we subtracted the mean visibility trend from each
individual visibility. The visibilities had been calibrated via
intermittent flux reference observations. We attribute the
  residual flux density dips/excesses to variations in the intrinsice
  flux density of SgrA*.  The combined light curve from all
telescopes is shown in Fig.~\ref{fig:1}.  
Here we can combine the data from different frequencies under the 
assumption that the spectral index of SgrA* does not change 
significantly during the flux density variations between 86 and 250~GHz.
In this case the flux density variations are frequency independent.
The light curve in Fig.~\ref{fig:1} shows two peaks, on
May 15 and 17 (There is a weaker, third possible peak on May 19).  
In Fig.~\ref{fig:1} we also show the daily flux
density averages of the 7\,mm VLBA observations that were conducted in
parallel.  The VLBA data follow the overall shape of the combined
CARMA/ATCA/30m lightcurve very well. In Fig.~\ref{fig:2} we show
residual maps from the four individual  tracks obtained with the CARMA
array. These maps were computed by subtracting the mean of all 4 maps
from the maps of the individual epochs.  This procedure clearly
reveals the excess flux density detected on May 17.

The first NIR flare detected during the multi-wavelength campaign (see
also Eckart et al.\ 2008) preceeded the combined mm/sub-mm monitoring
and the first maximum detected therein. The second NIR flare had overlap
with the CARMA observations, which show a 0.5~Jy 3mm flare that is
delayed with respect to the NIR flare.  In
Fig.~\ref{fig:3} we show both the VLT NIR and CARMA mm-lightcurve for
May 17.  The comparison reveals that the NIR flares occured during or
just before times of excess mm/sub-mm flux density.  For May 15 the
details of the NIR flare have been described by Eckart et al.\ (2008).
The observed time difference between the NIR and mm/sub-mm flares can
be interpreted in the framework of (adiabatic) expansion of jet or
disk synchrotron components - in full support of previous evidence for
adiabatic expansion (Eckart et al.\ 2006a, Yusef-Zadeh et al.\ 2006).
Within the current radio size limits the apparent size of an expanding
jet must be small or foreshortened (240 $\mu$as at 43~GHz; Bower et
al. 2004).  However, a plasma component expanding within a
relativistic orbit or polarized quasi-periodic emission due to a
helical jet structure (Eckart et al.\ 2004, 2005, 2006a) cannot be
fully excluded yet.

Future measurements will concentrate on monitoring the flux density
variability of SgrA* in coordinated campaigns (radio/mm/MIR/X-ray)
and in polarized radio/NIR emission.  NIR telescopes with large
aperturtes (VLT, Keck, LBT, TNT) will be best suited to separate
SgrA* especially during faint phases from the surrounding high
velocity stars.  Similarly, mm-interferometers like the PdBI, CARMA,
ATCA, and, in future, ALMA can separate SgrA* from the thermal emission
of the CND and the mini-spiral.  Therefore, the combination of these
observing facilities will allow us to study the evolution of expanding
synchrotron components.

With near future mm-VLBI at frequencies at and above 230 GHz, the 
imaging of the central region of SgrA* will become possible, allowing 
to map out the structure of SgrA* on spatial scales of only a
few gravitational radii (Doeleman et al.\  2008). 
By this it should be possible to directly test the hypothesis of spiraling plasmons or density waves in the accretion disk.

\section*{Acknowledgements}
Part of this work was supported by the German
\emph{Deut\-sche For\-schungs\-ge\-mein\-schaft, DFG\/} via grant SFB 494.
L. Meyer, K. Muzic, M. Zamaninasab, D. Kunneriath, and R.-S. Lu,
 are members of the International Max Planck Research School (IMPRS) for 
Astronomy and Astrophysics at the MPIfR and the Universities of 
Bonn and Cologne. RS acknowledges support by the Ram\'on y Cajal
programme by the Ministerio de Ciencia y Innovaci\'on of the
government of Spain.

\end{document}